\documentclass[aps,twocolumn,prb,10pt,showpacs]{revtex4}
\usepackage{epsfig}
\begin{document}

\title{Spin polarization and magneto-luminescence of confined electron-hole
systems}
\author{Boris Anghelo Rodriguez$^1$}
\email{banghelo@fisica.udea.edu.co}
\author{Augusto Gonzalez$^{1,2}$}
\email{agonzale@fisica.udea.edu.co}
\affiliation{$^1$Departamento de F\'{\i}sica, Universidad de Antioquia,
AA 1226, Medell\'{\i}n, Colombia\\ 
$^2$Instituto de Cibern\'etica, Matem\'atica y F\'{\i}sica, Calle E 
309, Vedado, Habana 4, Cuba}

\date{\today}

\begin{abstract}
A BCS-like variational wave-function, which is exact in the infinite field limit, 
is used to study the interplay among Zeeman energies, lateral confinement and particle
correlations induced by the Coulomb interactions in strongly pumped neutral quantum
dots. Band mixing effects are partially incorporated by means of field-dependent masses and g
factors. The spin polarization and the magneto-luminescence are computed as functions of the 
number of electron-hole pairs present in the dot and the applied magnetic field.
\end{abstract}
\vspace{.5cm}
\maketitle

\section{Introduction}

In the present paper, we study the spin polarization and the
magneto-luminescence of a neutral, medium-size quantum dot (qdot) subjected to a strong
(pulsed) laser pumping and a strong magnetic field. There are many good reasons to study the
properties of this system.

On one hand, recent magneto-tunneling experiments \cite{oosterkamp,haw} have stated very
clearly the role of Zeeman energies, lateral confinement and Coulomb repulsion in the spin
polarization of a qdot filled with $N$ electrons. At ``filling factors'' 
$2> \nu \agt 1$, ground-state rearrangements lead to significant oscillations of the
conductance peak positions as a function of the magnetic field. The situation could be even
more interesting for electron-hole (e-h) clusters, where additional e-h correlations come
into play. Traces of spin rearrangements shall be seen in the luminescence and other optical
properties. 

On the other hand, extensive measurements of quantum well (qwell) luminescence exist for
different
e-h densities (laser excitation power) and polarizations \cite{lumin}. A BCS-based theory has
been proposed \cite{tejedor}. Very high magnetic fields, up to 60 T, have been applied mainly
to low-density systems, and the results have been interpreted in terms of isolated neutral
and negatively charged excitons \cite{kim1}. Analogous experiments in qdots are lacking, but
the available experimental resources are enough to create a high population of excitons in a
medium-size qdot in a strong magnetic field. 6-exciton luminescence has been undoubtedly
observed in single small InAs quantum dots at $B=0$ \cite{gershoni}. In qwells, very high e-h
densities ($10^{13}$ pairs/cm$^2$) have been achieved with pulsed pumping \cite{kim2}.

Our model quantum dot is made up from a symmetrical GaAs-AlGaAs quantum well, 8.5 nm wide
in the growth direction. Stress is supposed to induce a lateral confinement, which
is described by a parabolic potential with $\hbar \Omega \approx 1.2$ meV.
An electron-hole system is created in the dot by a strong 5 ps pulsed laser, as it is 
experimentally done for example in Ref.\onlinecite{kim2}. The mean lifetime of this system is
around hundred of ps or even longer, i. e. a time scale much greater than the
characteristic times ($\approx$ 1 ps) to reach equilibrium \cite{time}. Thus, at
very low temperatures we end up with an ``stable'' $N$-pair e-h cluster in its
ground state.

Results for spin-up and -down densities, hole and electron spin polarizations and for the
position and magnitude of the coherent luminescence peak as functions of the magnetic field
and of the mean number of excitons are presented below. The theoretical framework used is a
BCS-like wave function, corrected against particle number non-conservation by means of a
Lipkin-Nogami procedure \cite{LN,yo}. This wave function is able to reproduce the expected
``Bose condensed'' state in the $B \to \infty$ limit \cite{lerner,dzyubenko,paquet}. A big basis of 
one-particle states is used, which includes up to 3 Landau levels and 202 states per
Landau level. We consider systems with up to 40 e-h pairs.

The plan of the paper is as follows. In Section II, the model to be employed is described in
details. The basics of our theoretical scheme are summarized in Section III. In Section IV,
the main results are presented and discussed. Concluding remarks are given in the last
section.

\section{The Model}

We consider a system of $N$ electrons and $N$ holes confined in a quasi two-dimensional
quantum dot, and in the presence of a perpendicular magnetic field. As mentioned above, the
qdot is made up from a 85 {\AA}-wide symmetric qwell. A parabolic potential confines the
motion of the particles in the plane perpendicular to the grow direction. The first qwell
sub-band approximation is used, i. e. the confining energies along the $z$ direction are
written as
$E_{e(h)}^{z}=\frac{\hbar^{2}\pi^{2}}{2m_{e(h)}L_{z}^{2}}k_{e(h)}^{2}$, with $k_{e(h)}=1$.
Notice that for the $85$ {\AA}-width GaAs well, the gap to the second
qwell state is 
$\Delta E_e^z\approx 210$ meV, and $\Delta E_h^z\approx 39$ meV for
electrons and holes respectively, whereas the typical Coulomb energy is $E_{Coulomb}
\approx 3.18 \sqrt{B[T]}$ meV. 
This model is a common theoretical framework in the study of strained or
self-assembled quantum dots \cite{qdot}. By using the symmetric gauge,
$\vec{A}(\vec{r})=\frac{\vec{B}\times \vec{r}}{2}$, 
dimensionless coordinates, $\vec{r}\rightarrow l_B\vec{r}$, with
$l_B=\sqrt{2\hbar c/eB}$, and using Landau level (LL) 
states $\left\{ |i\rangle =|n_i,m_i,s_i\rangle \right\} $ ($ n_i=0,1,\cdots
;\;\;m_i=-\infty ,\cdots ,-1,0,1,\cdots ,\infty;$ and $s_i=\pm 1/2$ are the
radial, angular momentum and spin quantum numbers, respectively) as set of
one-particle states, we can write the Hamiltonian in second quantization as:

\begin{eqnarray}
H&=& \sum_i\left\{ \frac{\hbar \omega _c^e}2\varepsilon _i^e + E_{e_{i}}^{Zeeman}\right\} 
e_i^{\dag}e_i \nonumber\\
&+&\sum_i\left\{ \frac{\hbar \omega _c^h}2\varepsilon _i^h + E_{h_{i}}^{Zeeman}\right\} 
h_i^{\dag}h_i \nonumber \\
&+& \sum_{ij} \hbar \langle i|\vec{r}^2|j \rangle \left\{\frac{\Omega^2}{\omega _c^e}
e_i^{\dag}e_j + \frac{\Omega^2}{\omega _c^h}
h_i^{\dag}h_j \right\} \nonumber\\
&+& N\left\{E_{gap}+E_e^z(k_e=1)+E_h^z(k_h=1) \right\} \nonumber\\
&+& \frac{e^2}{\epsilon l_B}\left\{ \frac 12\sum_{ijkl}\langle ij|\frac 1{
\left| \vec{r}\right| }|kl\rangle e_i^{\dag
}e_j^{\dag }e_le_k\right.\nonumber\\
&+&\frac 12
\sum_{ijkl}\langle ij|\frac 1{\left| \vec{r}\right| }|kl\rangle h_i^{\dag}h_j^{\dag }
h_lh_k \nonumber\\
&-&\left.\sum_{ijkl}\langle ij|\frac
1{\left| \vec{r}\right| }
|kl\rangle e_i^{\dag }h_j^{\dag }h_le_k\right\}, 
\label{hamiltoniano}
\end{eqnarray}

\noindent
where $m_e(m_h)$ are the electron and hole effective masses, $\Omega$ is the dot confining 
frequency, and $\epsilon $ is
the dielectric constant. $E_{gap}$ is the gap separation between the conduction and
valence bands, $E^{Zeeman}_{e_i(h_i)}=\pm g_{e(h)}\mu B s_i^{e(h)}$,
with $g_{e(h)}$ the effective g-factors, $\mu $ is the Bohr magneton,
$s_i^{e(h)}$ are the $z$-components of the ith particle spin,
$\varepsilon _i^{e(h)}=2n_i+|m_i|\pm m_i+1$ are the LL energies for
electrons (holes), $\omega_c^{e(h)}=eB/m_{e(h)}c$ is the electron (hole) cyclotron frequency, 
and $e_i(h_i),\;e_i^{\dag }(h_i^{\dag })$ are the electron (hole) destruction and
creation operators. Conventionally, we write $s_h=\pm 1/2$ for the two branches of the heavy
hole sub-band. To the electronic $j_z=-3/2$, for example, we ascribe $s_h=1/2$.

The effective parameters entering the Hamiltonian (masses and g factors) are 
magnetic field- and width-dependent magnitudes to approximately account for band mixing.  
For the $85$ {\AA}  well, we fitted the experimental in-plane heavy hole
mass \cite{mh},
thus obtaining:

\begin{equation}
 m_h^{85 \mbox{\AA}}(B)= \left\{
\begin{array}{ll}
0.17,  & \mbox{for $B<10$ T} \\ 
\frac{0.17+0.0168 B}{1+0.023 B}, & \mbox{for $B>10$ T}.
\end{array}
 \right.
\end{equation}

Experimentally, the dependence of the electron g-factor on well width and magnetic
field is well determined \cite{ge5,ge1,ge2,ge3,ge4}, and in our case we have:

\begin{equation}
g_{e}^{85\mbox{\AA}}(B)=-0.1667+0.0052 \;B[\rm{T}].
\label{eq3}
\end{equation}

\noindent
The dependence of the hole g-factor on the magnetic field for high $B$ values, however, is
not properly determined \cite{ge2,ge3}. Here we assume a linear behavior with the field, as
in the electron case, and fitted it to \hspace{4cm} \cite{kim1}. The result is

\begin{equation}
g_{h}^{85\mbox{\AA}}(B)=-0.05 \;B[\rm{T}].
\end{equation}

\noindent
Notice that in the $B\to 0$ limit, $g_h$ vanish because the exciton
(X) g-factor, defined as $g_X=g_e+g_h$, and the electron g-factor are equal
in the $85$ {\AA} width well\cite{ge2}.

With this parameterization, $g_e$ changes sign at $B_c\approx 32$ T. Consequences of this
facts are discussed below.

\section{The BCS scheme}

\begin{figure}
\begin{center}
\hspace{0.2truecm}
\psfig{file=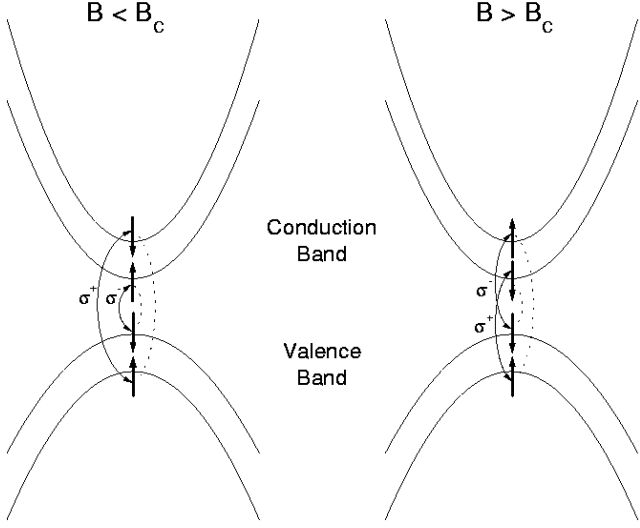,width=0.8\linewidth,angle=270} 
\end{center}
\caption{Schematics of the spin sub-band structure for the two field regimes. The dotted
lines show the paired electron and hole states in the BCS wave function. The interband
optical transitions for the two circular polarizations of light $\sigma^{\pm}$ are
indicated.}
\label{fig1}
\end{figure}

The BCS approach has been successfully applied by Paquet {\it et. al.} 
in the study of the two-dimensional (2D) e-h system \cite{paquet}, and by Fern\'andez-Rossier 
and Tejedor to the exciton gas in a qwell\cite{tejedor}. We used it previously in the
study of finite e-h systems at zero magnetic field \cite {yo}. We employ the
Lipkin-Nogami (LN) scheme to avoid particle number non-conservation in a
finite system. 
The BCS wave function is given by 

\begin{equation}
|BCS\rangle_{N}= \prod_{i} \left( u_{i}+v_{i}e_{i}^{\dag}
 h_{\bar{\imath}}^{\dag}\right)|0\rangle_{h}|0\rangle_{e}~~,
\end{equation}

\noindent
where $|0\rangle_{h}(|0\rangle_{e})$ are the respective electron (hole)
vacuum. The subscript ``$N$'' in the BCS function means that the average number of
pairs is just $N$, i.e. $\langle \hat{N} \rangle  = 
\langle BCS|\hat{N} |BCS \rangle _{N} = N$. $u_ {i}$ and $v_ {i}$ are the variational
parameters. They fulfill the normalization
conditions $u_ {i}^2+v_ {i}^2=1, \; \forall i$. The hole state $| \bar{\imath}
\rangle$ which is paired with the electron state $|i
\rangle=|n_{i},m_{i},s_{i} \rangle$ is different for the two magnetic field
regimes, reflecting the ground state spin structure, as it is show in Fig. \ref{fig1}.
For $B<32$ T, electrons and holes with opposite spins are paired, then $|\bar{\imath}
\rangle=|n_{i},-m_{i},-s_{i} \rangle$. Whereas for $B>32$ T, the energy is minimized 
when the $|i\rangle$ electronic state is paired with
$|\bar{\imath} \rangle=|n_{i},-m_{i},s_{i} \rangle$.

The total angular momentum (projection onto the z axis), corresponding to $|BCS \rangle_{N}$ is 
zero because the angular momentum of each pair is zero. The total electron or hole spin,
however, depend on the populations of spin-up and down components.

The detailed description of the LN method can be found in
Refs. \cite{LN,yo}. For
completeness,  we sketch the main results. The LN estimate for the ground state energy is
given by:

\begin{equation}
E_{LN}=E_{BCS}-2\lambda _1(\langle \hat{N}\rangle-N)-\lambda_2(\langle \hat{N}^2\rangle-N^2),
\end{equation}

\noindent
where $E_{BCS}$ is the expectation value of the effective Hamiltonian $\hat{H}=
H-N\left\{E_{gap}-E_e^z(k_e=1)-E_h^z(k_h=1)\right\}$ in $|BCS \rangle_{N}$:

\begin{eqnarray}
E_{BCS} &=&\mbox{}_N\langle BCS|\hat{H}|BCS\rangle_N \nonumber\\
&=&\sum_i\left\{ \frac{\hbar \omega _c^e}2\varepsilon _i^e+\frac{\hbar
 \Omega^2}{\omega _c^e}\varepsilon _i^{OA}+g_e\mu Bs_i^e\right\} v_i^2 \nonumber \\
&+&\sum_{\bar{\imath}}\left\{ \frac{\hbar \omega _c^h}2\varepsilon _{\bar{
 \imath}}^h+\frac{\hbar \Omega^2}{\omega _c^h}\varepsilon _{\bar{\imath}
 }^{OA}-g_h\mu Bs_{\bar{\imath}}^h\right\} v_i^2  \nonumber \\
&-&\frac{e^2}{\epsilon l_B}\left\{ \sum_{i\neq j}\langle ij|\frac 1{\left| 
 \vec{r}\right| }|ji\rangle \left( v_i^2v_j^2+v_iu_iv_ju_j\right) \right. \nonumber\\
&-&\left.\sum_i\langle ii|\frac 1{\left| \vec{r}\right| }|ii\rangle v_i^2\right\}.
\end{eqnarray}

\noindent
Notice that in the second term, the sum runs over hole states $|\bar{\imath}
\rangle$. $\varepsilon_{i}^{e}=\varepsilon_{\bar{\imath}}^{h}=
2n_{i}+|m_{i}|+m_{i}+1$ are the one-particle energies, and
$\varepsilon_{i}^{OA}=\varepsilon_{\bar{\imath}}^{OA}=
2n_{i}+|m_{i}|+1=\langle i|\vec{r}\,^2|i\rangle$ is the expectation value of the harmonic
potential. The mean value of the number of pairs is 

\begin{eqnarray}
N&=&\langle \sum_{i} e_{i}^{\dag}e_{i} \rangle_{BCS}=\langle \sum_{i}
h_{i}^{\dag}h_{i} \rangle_{BCS}  \nonumber \\
&=& \sum_{i}v_{i}^{2}.
\end{eqnarray}

The extrema conditions can be written in the standard form of gap equations

\begin{equation}
\Delta_{i}=\frac{e^2}{\epsilon l_{B}} \sum_{j(j\neq i)}\langle ij| \frac{1}
{|\vec{r}|}|ji\rangle \frac{\Delta_{j}}{2\sqrt{\Delta_{j}^{2}-
(\epsilon_{i}^{HF}-\mu)^{2}}},
\end{equation}

\noindent
where the Hartree-Fock energies are given by

\begin{eqnarray}
\epsilon _i^{HF} &=&\frac 14(\hbar \omega _c^e+\omega _c^h)(2n_i+|m_i|+m_i+1) 
\nonumber\\
&+&\frac 12\left( \frac{\hbar \Omega _e^2}{\omega _c^e}+\frac{\hbar \Omega
_h^2}{\omega _c^h}\right) (2n_i+|m_i|+1)  \nonumber \\
&+&\frac \mu 2B(g_es_i-g_hs_{\bar{\imath}})-\frac{e^2}{2\epsilon l_B}\langle
ii|\frac 1{\left| \vec{r}\right| }|ii\rangle\nonumber\\  
&-&\frac{e^2}{\epsilon l_B}\sum_{j(j\neq i)}\langle ij|\frac 1{\left| \vec{r}
\right| }|ji\rangle v_j^2-\lambda _2(N-v_i^2),
\end{eqnarray}

\noindent
and we have used the usual BCS parameterization

\begin{equation}
v_{i}^{2}=\frac{1}{2} \left(1-\frac{\epsilon_{i}^{HF}-\mu} {\sqrt
{\Delta_{i}^{2}-(\epsilon_{i}^{HF}-\mu)^{2}}} \right).
\end{equation}

\noindent
The chemical potential $\mu=\lambda_{1}+\lambda_{2}/2$ was introduced to
fix the particle number, and $\lambda_{2}$ is determined in the LN scheme as:

\begin{eqnarray}
\lambda_{2}&=&\left\{ \langle \hat{H}\hat{N}^{2} \rangle \left( N^{2}-
 \langle \hat{N}^{2} \rangle \right)+ \langle \hat{H}\hat{N} \rangle
 \left( \langle \hat{N}^{3} \rangle- \langle N \hat{N}^{2} \rangle \right) 
\right.\nonumber\\
&+&\left. \langle \hat{H} \rangle \left( \langle \hat{N}^{2} \rangle^{2}-
\langle N \hat{N}^{3} \rangle \right)\right\}
\left\{ \langle \hat{N}^{2} \rangle^{3}+ \langle \hat{N}^{3} \rangle^{2} 
\right.\nonumber\\
&+&\left. N^{2} \langle \hat{N}^{4} \rangle - \langle
\hat{N}^{2} \rangle \left( 2N\langle \hat{N}^{3} \rangle + \langle
\hat{N}^{4} \rangle \right)\right\}^{-1},
\end{eqnarray}

\noindent
where the expectation values $\langle \cdots \rangle $ are taken in the 
$|BCS\rangle _N$ state. The resulting equations are solved iteratively up to
a precision of $10^{-12}$ in $\epsilon _i^{HF}$. Calculations were performed
for $20\leq N\leq 40$ pairs and $606$ one-particle LL states.

\begin{figure}
\begin{center}
\psfig{file=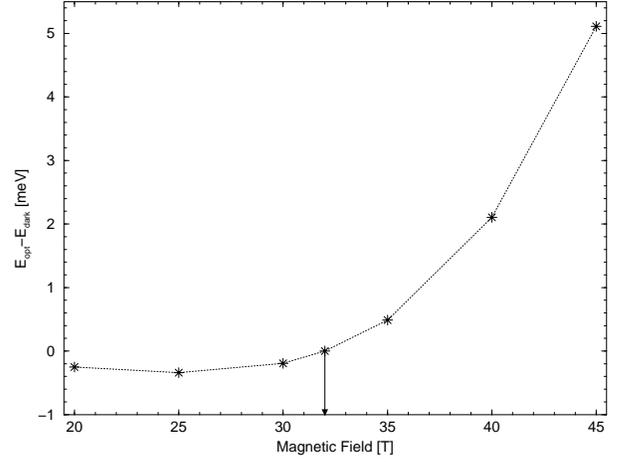,width=0.7\linewidth,angle=270} 
\end{center}
\caption{Dependence of  $E_{opt}-E_{dark}$ with the magnetic field. In figures 2 - 5,
results for the $N=40$ exciton system are presented.}
\label{fig2}
\end{figure}

\section{Polarization and magneto-luminescence}

As shown below, the spin polarization of the electronic subsystem increases as
the magnetic field is increased. At $B=32$ T, the electronic $g$-factor changes 
sign according to Eq. \ref{eq3}, leading to a rearrangement of electron-hole
pairing in the absolute ground state. The kind of pairing minimizing the
energy is represented in Fig. \ref{fig1} by dashed lines. For $B>32$ T, the
ground state is dark, and it is not even clear whether it can be reached
by means of light excitation followed by spin relaxation processes. Thus, 
besides the ground state, for $B>32$ T, we compute also the lowest BCS
state with $\sigma^+$ and $\sigma^-$ excitons. Below, we present results for the $N=40$ system, 
obtained with three LLs and 202 states per level, i. e. a total of 606 one-particle states. 

As mentioned above, two BCS functions may be constructed. One in which 
optical excitons are formed, and a second one in which dark excitons are
formed. The difference $E_{opt}-E_{dark}$ is drawn in Fig. \ref{fig2} 
as a function of $B$, showing that the dark state becomes the ground
state when the value $B\approx 32$ T is crossed and the electronic
sub-bands are re-ordered. The absence of efficient spin relaxation
mechanisms may, however, prevent the actual ground state to be occupied.    

\begin{figure}
\begin{center}
\psfig{file=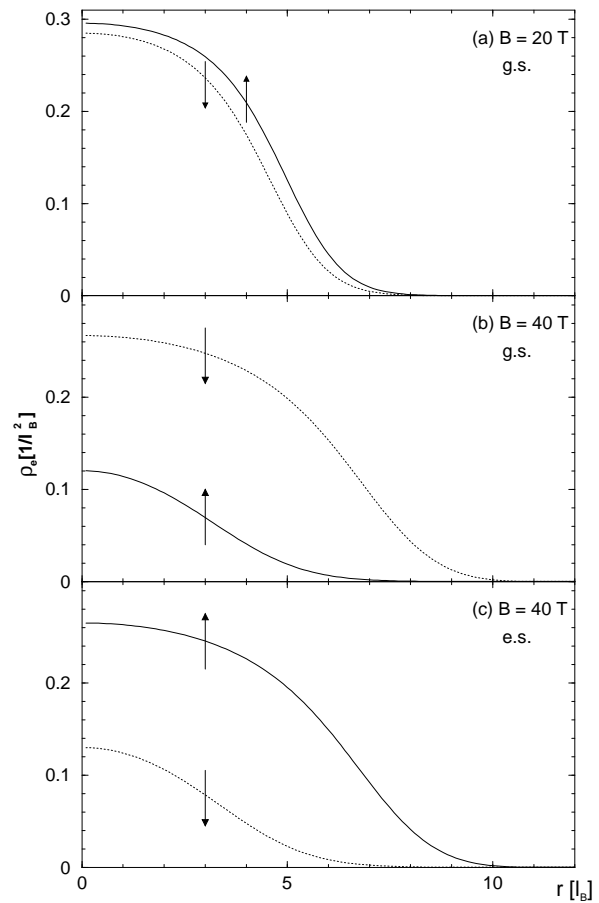,width=0.9\linewidth} 
\end{center}
\caption{Spin-up and down densities. (a) Ground state at $B=20$ T,
(b) Ground (dark) state at $B=40$ T, and (c) Excited (optically 
active) state at $B=40$ T.}
\label{fig3}
\end{figure}

It is interesting to note that the difference $E_{opt}-E_{dark}$, in
the magnetic field interval shown in Fig. \ref{fig2} is very close to the difference
between the electronic Zeeman energies. A simple qualitative picture can be 
offered for the understanding of this and the next figures. The properties
of the system are roughly determined by the holes because of the 
competition among the hole Zeeman energies and the total harmonic and 
Coulomb energies. The hole occupations are thus very similar for 
optical and dark states. The form of e-h pairing provides the ``fine
structure''. That is, the minimization of the energy leads to a definite 
pairing. 

Scaled spin-up and -down densities, obtained from

\begin{eqnarray}
\rho_e^{(\uparrow,\, \downarrow)}(\vec{r}) &=& 
\sum_{i=\{n_i,m_i,(\uparrow,\, \downarrow)\} \atop j=\{n_j,m_j,(\uparrow,\, \downarrow)\}}
 \phi_i^e(\vec{r}) \phi_j^e(\vec{r})\: \mbox{}_N\langle BCS|e_i^{\dag}e_j|BCS\rangle_N
\nonumber \\
&=& \frac{e^{-r^2}}{\pi} \sum_{i=\{n_i,m_i,(\uparrow, \downarrow)\}} \frac{n_i!\,
r^{2|m_i|}\, v_i^2} {(n_i+|m_i|)!}\left[ L_{n_i}^{|m_i|} (r^2) \right]^2,\nonumber \\
\end{eqnarray}

\noindent
are shown in Fig. \ref{fig3}. At $B=40$ T, we show the ground (dark) and 
optically-active excited-state densities, which are almost inverted in
agreement with the argument given above. Notice that there are only four
excess spin-up electrons at $B=20$ T. These small net polarizations for
high magnetic fields are related to the attractive character of the e-h
interaction. Unlike pure electron systems, small-radius orbits maximize
e-h attraction, and the competition between Zeeman and Coulomb energies
starts at higher fields.

\begin{figure}
\begin{center}
\psfig{file=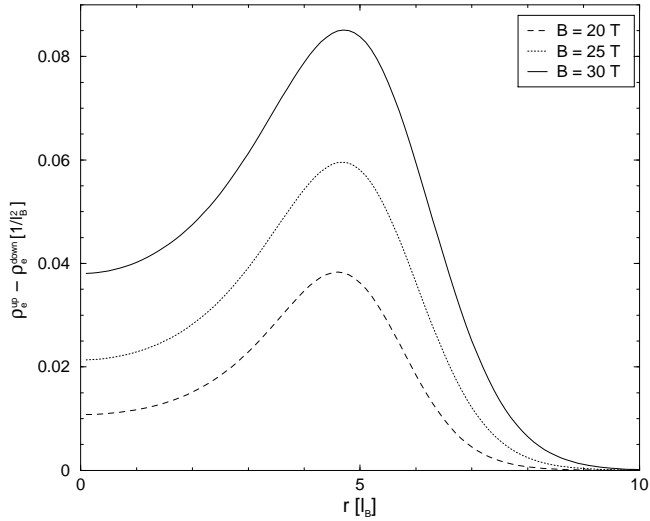,width=0.8\linewidth,angle=-90} 
\end{center}
\caption{Difference between ground-state spin-up and -down densities
for various magnetic field values.}
\label{fig4}
\end{figure}

The ``hole dominance picture'' leads to changes in the polarization, as
$B$ is increased, through the reconstruction of the droplet edge, in a 
way very similar to electrons near filling factor one \cite{edge}. This
fact is illustrated in Fig. \ref{fig4}, where the difference between 
ground-state spin-up and -down densities for different values of $B$ are shown. 
It is evident that polarized densities differ mainly at the edge, and
that changes in the polarization are more significant at the edge.

\begin{figure}
\begin{center}
\psfig{file=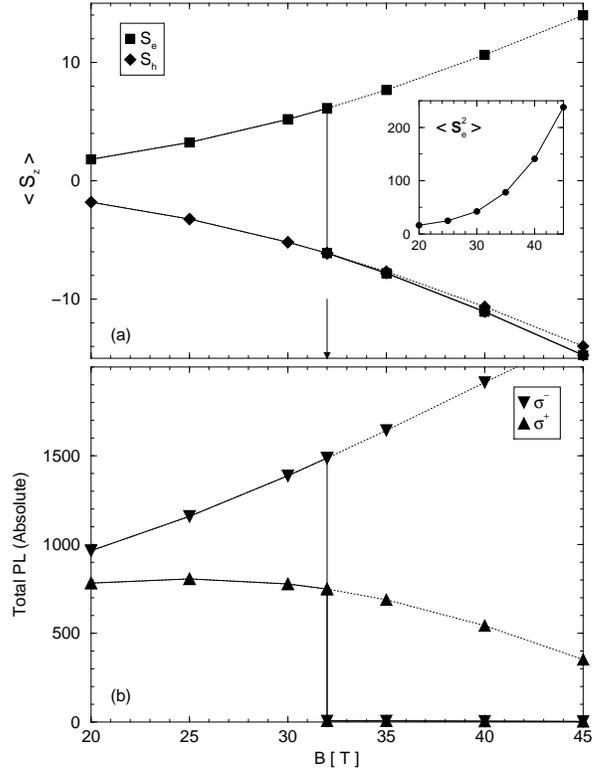,width=0.9\linewidth} 
\end{center}
\caption{(a) Electron and hole spin polarizations. For $B>32$ T, 
both ground-state (solid line) and excited-state (dashed line)
properties are drawn. The inset shows the total electron spin
squared. (b) Luminescence intensities for both $\sigma^-$ and 
$\sigma^+$ polarizations.}
\label{fig5}
\end{figure}

\begin{figure}
\begin{center}
\psfig{file=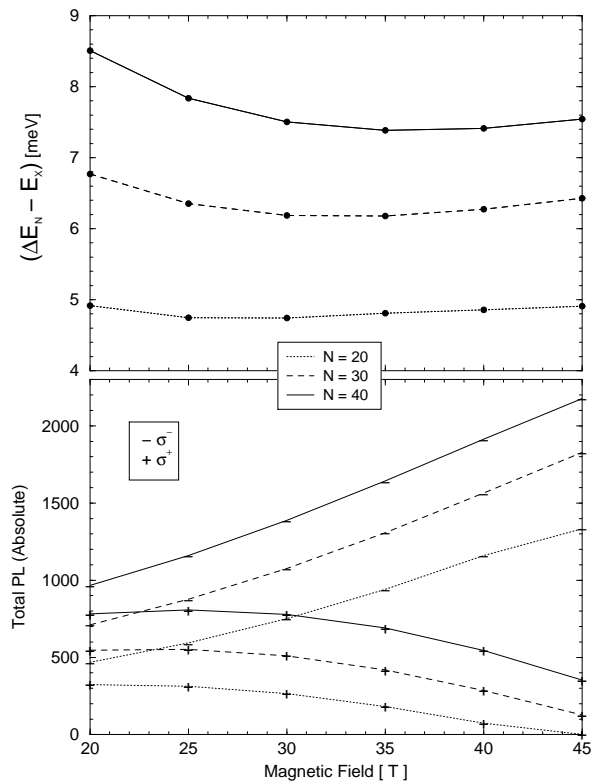,width=0.9\linewidth} 
\end{center}
\caption{Dependence of the position of the luminescence line (see main
text) and the intensities on the number of pairs in the dot. $E_X$ is 
the $\sigma^-$ exciton energy.}
\label{fig6}
\end{figure}

The following two figures, Figs. \ref{fig5} and \ref{fig6}, contain 
the main results of our paper. Electron and hole spin polarizations
are drawn in Fig. \ref{fig5}a as a function of $B$. Solid lines
refer to the ground state, while the dashed lines for $B>32$ T 
refer to the optically active state. Notice that even at a high
field value like 45 T, the electronic polarization is only 70 \%.
Notice also the change in sign of the ground-state electronic
spin at $B=B_c$. The inset shows the total ground-state spin
squared, computed from 

\begin{equation}
\langle \vec{S}_e^2 \rangle_{BCS} = \frac{3}{4}N 
+\sum_{i,j} s_i s_j v_i^2 v_j^2-\sum_i s_i v_i^4+
\frac{1}{2}\sum_{\alpha} v_{\alpha \uparrow}^2 v_{\alpha \downarrow}^2.
\end{equation}

The total (coherent) magnetoluminescence intensity for both
$\sigma^+$ and $\sigma^-$ polarizations is presented in Fig.
\ref{fig5}b. We compute it for $\sigma^-$ polarization, for example,
from the expression

\begin{eqnarray}
I_{total}^{\sigma^-}&=&
\sum_f \Bigl| \,\mbox{}_{N-1} \langle f|P_{\sigma^{-}}|BCS \rangle_N\, \Bigr|^2 
\nonumber\\
&\propto&
\mbox{}_N \langle BCS|P_{\sigma^{-}}^{\dagger}\,P_{\sigma^{-}}|BCS \rangle_N \nonumber \\
&=& \left\{
\begin{array}{ll}
 \sum_{\alpha} v_{\alpha \uparrow}^4 + 
  \left( \sum_{\alpha} u_{\alpha \uparrow}v_{\alpha \uparrow} \right)^2, & 
  \mbox{bright state} \\
 \sum_{\alpha} v_{\alpha \uparrow}^2 v_{\alpha \downarrow}^2, & \mbox{dark state},
\label{pltotal}
\end{array}
\right. 
\end{eqnarray}

\noindent
where $|f\rangle$ is a basis of $N-1$ particle states ,
$\alpha =(n,\,m)$ is a composed index and, 
$P_{\sigma^{-}}=\sum_{\alpha} e_{\alpha \uparrow}h_{\bar \alpha \downarrow}$ is the
interband dipole transition operator for the $\sigma^-$ circularly polarized light.
Notice that $I_{total}$ is the integrated luminescence, corresponding to the transition from
the given initial BCS state to any final state.
The convention for solid and dashed lines is the same as in Fig. \ref{fig5}a.
It shall be stressed that the degree of polarization, defined from

\begin{equation}
\frac{I_{total}^{\sigma^-}-I_{total}^{\sigma^+}}{I_{total}^{\sigma^-}+I_{total}^{\sigma^+}},
\label{eq16}
\end{equation}

\noindent
follows very well the behaviour of $\langle S_z \rangle$, i. e. the difference 
between the occupation of spin-up and -down sub-bands. This polarization is nearly
10 \% at 20 T, and around 70 \% at 45 T.  

Finally, in Fig. \ref{fig6} we show results for the position of the luminescence
line and the intensities as functions of the numbers of pairs in the dot. In our 
computations, the energy is corrected against non-conservation of the
total number
of excitons, $N$. Conservation of the number of $\sigma^+$ and $\sigma^-$ excitons
is not properly taken into account. Thus, we can not exactly compute the position
of the $\sigma^+$ and $\sigma^-$ lines. In place of it, we show in Fig. \ref{fig6}a
the difference $\Delta E_N=E_{LN}(N)-E_{LN}(N-1)$ relative to the exciton 
$\sigma^-$ line, $E_X$. The following interesting properties can be
noticed in this figure: a) A blueshift as the number of excitons is raised 
\cite{kim2}. It is around 0.2
meV/exciton at $N=20$, and 0.15 meV/exciton at $N=30$, and b) An apparent minimum of each 
curve at $B$ around 30 T, i. e. at $B_c$. On the other hand, the intensity
(Fig. \ref{fig6}b) shows an increase with $N$ for both polarizations, as one would expect
from coherent emission.

\section{Concluding remarks}

We have computed the spin polarization and the luminescence of a quantum dot in which a
mean number of electron-hole pairs, $N$, have been created by a laser pulse. Band
mixing effects were approximately taken into account by means of well width- and 
magnetic field-dependent masses and $g$-factors. For the model under study, the 
electron $g$-factor vanishes at $B_c\approx 32$ T. It means that, for magnetic field
values around $B_c$, the electron polarization, and thus the ratio of intensities 
given by formula (\ref{eq16}), is determined as a result of the interplay among 
Coulomb, confinement and hole Zeeman energies. The net polarization is only 70 \% at
$B=45$ T because of the attractive electron-hole interaction.

The general features found in our calculations, i. e. relatively small polarizations
even at high magnetic field values, blueshift of the luminescence lines with an
increase of the laser power, etc seem to be not related to the specific parametrization
used for carrier masses and $g$-factors. 

The developed computational scheme may be applied to many other interesting situations,
from which two of them may be distinguished. The first is the stationary regime, in which
constant populations of $\sigma^-$ and $\sigma^+$ excitons arise as a result of 
appropriate pumping, recombination and spin-flip processes \cite{lumin}. The second is the
study of the effects of hyperfine interactions between nuclear and electronic spins on
the position of the recombination lines, known
as Overhauser shifts \cite{overhauser}. Research along both directions is in progress.

\begin{acknowledgments}
The authors wish to thank R. P\'erez for helpful discussions. Support from the Colombian
Institute for Science and Technology (COLCIENCIAS), the Committee for Research of the 
University of Antioquia (CODI), and the Caribbean Network for Theoretical Physics are
gratefully acknowledged.
\end{acknowledgments}


\begin{thebibliography}{99}

\bibitem{oosterkamp} T. H. Oosterkamp, J. W. Janssen, L. P. Kouwenhoven, D. G. Austin, 
T. Honda, and S. Tarucha, Phys. Rev. Lett. {\bf 82,} 2931 (1999). 

\bibitem{haw} P. Hawrylak, C. Gould, A. Sachrajda, Y. Feng, and Z. Wasilewski,
Phys. Rev. B {\bf 59,} 2801 (1999).

\bibitem{lumin} T. C. Damen, L. Vi\~na, J. E. Cunningham, J. Shah, and L. J. Sham,
Phys. Rev. Lett. {\bf 67,} 3432 (1991).

\bibitem{tejedor} J. Fern\'andez-Rossier, and C. Tejedor, Phys. Rev. Lett. {\bf
78,} 4809 (1997).

\bibitem{kim1}  Y. Kim, F. M. Munteanu, C. H. Perry, D. G. Rickel, J. A.
Simmons, and J. L. Reno, Phys. Rev. B {\bf 61,} 4492 (2000); M. Hayne, C. L. Jones, R. Bogaerts,
C. Riva, A. Usher, F. M. Peeters, F. Herlach, V.V. Moshchalkov, and M. Henini, {\it ibid.} {\bf 59,}
2927 (1999). 

\bibitem{gershoni} E. Dekel, D. Gershoni, E. Ehrenfreund, D. Spektor, J. M. Garcia, and
P. M. Petroff, Phys. Rev. Lett. {\bf 80,} 4991 (1998); E. Dekel, D. V. Regelman, D. Gershoni, 
E. Ehrenfreund, W. V. Schoenfeld, and P. M. Petroff, cond-mat/0011166.
                                        
\bibitem{kim2} J. C. Kim and J. P. Wolfe, Phys. Rev. B {\bf 57,} 9861 (1998).

\bibitem{time} J. Shah, {\it Hot Carriers in Semiconductor Nanostructures}
(Academic, San Diego, 1992).

\bibitem{LN} J. Dobaczewski and W. Nozarewickz, Phys. Rev C {\bf 47,} 2418 (1993), and references 
cited therein.

\bibitem{yo} B. A. Rodr\'{\i}guez, A. Gonzalez, L. Quiroga, F. J. Rodriguez, and R. Capote, 
Int. J. Mod. Phys. B {\bf 14,} 71 (2000).

\bibitem{lerner} I. V. Lerner and Yu. E. Lozovik, Zh. Eksp. Teor. Fiz. {\bf 80,} 
1488 (1981) [Sov. Phys.{\bf--JEPT 53,} 763 (1981)].

\bibitem{dzyubenko} B. Dzyubenko and Yu. E. Lozovik, Fiz. Tverd. Tela. {\bf 25,}
1519 (1983). [Sov. Phys. Solid State {\bf 25,} 874 (1983)].

\bibitem{paquet} D. Paquet, T. M. Rice, and K. Ueda, Phys. Rev. B {\bf 32,} 
5208 (1985).

\bibitem{qdot} L. Jacak, P. Hawrylak, and A. Wojs, {\it Quantum Dots} (Springer-Verlag, 
Berlin, 1998).

\bibitem{mh}  B. E. Cole, J. M. Chamberlain, M. Henini, T. Cheng, W. Batty,
A. Wittlin, J. A. A. J. Perenboom, A. Ardavan, A. Polisski, and J.
Singleton, Phys. Rev B {\bf 55,} 2503 (1997).

\bibitem{ge5}  S. P. Najda, S. Takeyama, and N. Miura, Phys. Rev. B {\bf 40,}
6189 (1989).

\bibitem{ge1}  M.J. Snelling, G. P. Flinn, A.S. Plaut, R. T. Harley, A. C.
Trooper, R. Eccleston, and C. C. Phillips, Phys. Rev. B {\bf 44,} 11345
(1991).

\bibitem{ge2}  M. J. Snelling, E. Blackwood, C. J. McDonagh, R. T. Harley,
and C. T. B. Foxon, Phys. Rev. B {\bf 45,} 3922 (1992).

\bibitem{ge3}  N. J. Traynor, R. J. Waburton, M. J. Snelling, and R. T.
Harley, Phys. Rev. B {\bf 55,} 15701 (1997).

\bibitem{ge4}  M. Seck, M. Potemski, and P. Wyder, Phys. Rev. B {\bf 56,}
7422 (1997).

\bibitem{edge} C. de C. Chamon and X. G. Wen, Phys. Rev. B {\bf 49,} 8227 (1994).

\bibitem{overhauser} S. W. Brown, T. A. Kennedy, D. Gammon, and E. S. Snow, Phys Rev B
{\bf 54,} 17339 (1996); S. W. Brown, T. A. Kennedy, and D. Gammon, Solid State Nucl.
Magn. Reson. {\bf 11,} 49 (1998), and references cited therein.

\end{thebibliography}
\end{document}